\documentclass[twocolumn,preprintnumbers,epsfig,amsmath,amssymb]{revtex4}

\usepackage{graphicx}
\usepackage{dcolumn}
\usepackage{bm}

\begin{document}

\title{ Exact Analytical Solution of the Constrained Statistical Multifragmentation Model }  

\author{Kyrill A. Bugaev}

\affiliation{Bogolyubov Institute for Theoretical Physics,
Kiev, Ukraine\\
%
%
Lawrence Berkeley National Laboratory, Berkeley, CA 94720, USA
}

\date{\today}

\begin{abstract}
A novel powerful mathematical method is presented, which allows us to find an  analytical
solution of 
a simplified version of the  statistical multifragmentation model with the  
restriction that the largest fragment size cannot exceed the finite volume of the system. 
A complete analysis of the isobaric partition singularities is done for finite system volumes. The finite
size effects for large fragments and the role of metastable (unstable) states are discussed. 

\noindent
\hspace*{0.0cm}{\bf PACS} numbers: 25.70. Pq, 21.65.+f, 24.10. Pa
\end{abstract}

\maketitle



\section{Introduction}

Exactly solvable models with phase transitions play a special role in the statistical
physics - they are the benchmarks of our understanding of critical phenomena that occur  in 
more complicated  substances.
They are our theoretical laboratories, where we can study the most fundamental problems of critical phenomena
which cannot be studied 
elsewhere. 
A great deal of 
progress was recently achieved  in our understanding of the multifragmentation phenomenon 
\cite{Bondorf:95, Gross:97, Moretto:97} when an exact analytical solution of a simplified version of 
the statistical multifragmentation model (SMM) \cite{Gupta:98,Gupta:99} 
was found in Refs. \cite{Bugaev:00,Bugaev:01}. 
This exact solution allowed us to elucidate the role of the Fisher exponent $\tau$ on 
the properties of (tri)critical point and to show explicitly \cite{Reuter:01}
that in the SMM the relations between $\tau$
and other critical indices 
differ from the corresponding relations of a well known Fisher droplet model \cite{Fisher:64}. 
Note that these questions {\it in principle} cannot be clarified either  within  the widely used 
mean-filed approach or numerically.   

Despite this success, the application of the exact analytical  solution \cite{Bugaev:00,Bugaev:01}
to the description of experimental data is very limited because this solution  
corresponds to an infinite system volume.
To extend the formalism for finite volumes it is  
necessary to account for the finite size  and geometrical shape of 
the largest fragments, if  they  are comparable with the system volume.  
It is clear that these corrections may be important for not too dilute
systems. 
Therefore, to have a more realistic model
it is necessary  to abandon the arbitrary size of largest fragment and 
consider the constrained SMM (CSMM) in which
the largest fragment  size 
is explicitly related to the volume $V$ of the system. 
This will allow us to 
solve the CSMM analytically at finite volumes, to consider
how the first order phase transition 
develops from the singularities of the isobaric partition \cite{Goren:81} in thermodynamic limit, 
and to  
study the finite size effects for  large fragments.


\section{Laplace-Fourier Transformation}

The system states in the SMM are specified by the multiplicity
sets  $\{n_k\}$
($n_k=0,1,2,...$) of $k$-nucleon fragments.
The partition function of a single fragment with $k$ nucleons is
\cite{Bondorf:95}:
$
V \phi_k (T) = V\left(m T k/2\pi\right)^{3/2}~z_k~
$,
where $k=1,2,...,A$ ($A$ is the total number of nucleons
in the system), $V$ and $T$ are, respectively, the  volume
and the temperature of the system,
$m$ is the nucleon mass.
The first two factors  on the right hand side (r.h.s.) 
of 
the single fragment partition 
originate from the non-relativistic thermal motion
and the last factor,
 $z_k$, represents the intrinsic partition function of the
$k$- nucleon fragment. Therefore, the function $\phi_k (T)$ is a phase space
density of the k-nucleon fragment. 
For \mbox{$k=1$} (nucleon) we take $z_1=4$
(4 internal spin-isospin states)
and for fragments with $k>1$ we use the expression motivated by the
liquid drop model (see details in \mbox{Ref. \cite{Bondorf:95}):}
$
z_k=\exp(-f_k/T),
$ with fragment free energy
\begin{equation}\label{one}
f_k = - W(T)~k 
+ \sigma (T)~ k^{2/3}+(\tau + 3/2) T\ln k~,
\end{equation}
with $W(T) = W_{\rm o} + T^2/\epsilon_{\rm o}$.
Here $W_{\rm o}=16$~MeV is the bulk binding energy per nucleon.
$T^2/\epsilon_{\rm o}$ is the contribution of
the excited states taken in the Fermi-gas
approximation ($\epsilon_{\rm o}=16$~MeV). $\sigma (T)$ is the
temperature dependent surface tension parameterized
in the following relation:
$
\sigma (T)=\sigma_{\rm o}
[(T_c^2~-~T^2)/(T_c^2~+~T^2)]^{5/4},
$
with $\sigma_{\rm o}=18$~MeV and $T_c=18$~MeV ($\sigma=0$
at $T \ge T_c$). The last contribution in Eq.~(\ref{one}) involves the famous Fisher's term with
dimensionless parameter
$\tau$. 

The canonical partition function (CPF) of nuclear
fragments in the SMM
has the following form:
\begin{equation} \label{two}
\hspace*{-0.2cm}Z^{id}_A(V,T)=\sum_{\{n_k\}} \left[\prod_{k=1}^{A}\frac{\left[V~\phi_k(T) \right]^{n_k}}{n_k!} \right]
{\textstyle \delta(A-\sum_k kn_k)}\,.
\end{equation}
In Eq. (\ref{two}) the nuclear fragments are treated as point-like objects.
However, these fragments have non-zero proper volumes and
they should not overlap
in the coordinate space.
In the excluded volume (Van der
Waals) approximation
this is achieved
by substituting
the total volume $V$
in Eq. (\ref{two}) by the free (available) volume
$V_f\equiv V-b\sum_k kn_k$, where
$b=1/\rho_{{\rm o}}$
($\rho_{{\rm o}}=0.16$~fm$^{-3}$ is the normal nuclear density).
Therefore, the corrected CPF becomes:
$
Z_A(V,T)=Z^{id}_A(V-bA,T)
$.
The SMM defined by Eq. (\ref{two})
was studied numerically in Refs. \cite{Gupta:98,Gupta:99}.
This is a simplified version of the SMM, e.g. the symmetry and
Coulomb contributions are neglected.
However, its investigation
appears to be of  principal importance
for studies of the liquid-gas phase transition.

The calculation of $Z_A(V,T)$  
is difficult because of the constraint $\sum_k kn_k =A$.
This difficulty can be partly avoided by calculating the grand canonical
partition function:
\begin{equation}\label{three}
{\cal Z}(V,T,\mu)~\equiv~\sum_{A=0}^{\infty}
\exp\left({\textstyle \frac{\mu A}{T} }\right)
Z_A(V,T)~\Theta (V-bA) ~,
\end{equation}
where $\mu$ denotes a chemical potential.
The calculation of ${\cal Z}$  is still rather
difficult. The summation over $\{n_k\}$ sets
in $Z_A$ cannot be performed analytically because of
additional $A$-dependence
in the free volume $V_f$ and the restriction
$V_f>0 $.
This problem was resolved   \cite{Bugaev:00,Bugaev:01} 
by the Laplace transformation method to 
the so-called
isobaric ensemble \cite{Goren:81}.

In this work  we would like to consider a more strict constraint
$\sum\limits_k^{\alpha V/b} k~n_k =A$, where the size
of the largest fragment cannot exceed the total volume of the system 
(the parameter $\alpha$  is introduced for convenience). 
A similar restriction should be also applied to the upper limit of the product in 
all partitions $Z_A^{id} (V,T)$, $Z_A(V,T)$
and ${\cal Z}(V,T,\mu)$ introduced above 
(how to deal with the real values of $\alpha V/b$, see later). 
Then the  model with this constraint, the CSMM,  cannot be solved by the Laplace 
transform method, because the volume integrals cannot be evaluated due to a complicated 
functional $V$-dependence.  
However, the CSMM can be solved analytically with the help of  the following identity 
\begin{equation}\label{four}
G (V) = 
%
\int\limits_{-\infty}^{+\infty} d \xi~ \int\limits_{-\infty}^{+\infty}
  \frac{d \eta}{\sqrt{2 \pi}} ~ 
{\textstyle e^{ i \eta (V - \xi) } } ~ G(\xi)\,, 
\end{equation}
which is based on the Fourier representation of the Dirac $\delta$-function. 
The representation (\ref{four}) allows us to decouple the additional volume
dependence and reduce it to the exponential one,
which can be dealt by the usual Laplace transformation.
Indeed,
with the help of (\ref{four})   
the Laplace transform of the CSMM grand canonical partition (GCP) (\ref{three})  
can be done analytically: 
\begin{eqnarray} \label{five}
&&\hat{\cal Z}(\lambda,T,\mu)~\equiv ~\int_0^{\infty}dV~{\textstyle e^{-\lambda V}}
~{\cal Z}(V,T,\mu) = \nonumber\\
&&\hspace*{-0.1cm}\int_0^{\infty}\hspace*{-0.2cm}dV^{\prime}
\int\limits_{-\infty}^{+\infty} d \xi~ \int\limits_{-\infty}^{+\infty}
\frac{d \eta}{\sqrt{2 \pi}} ~ { \textstyle e^{ i \eta (V^\prime - \xi) - \lambda V^{\prime} } } \times 
\nonumber \\
&& \sum_{\{n_k\}}\hspace*{-0.1cm} \left[\prod_{k=1}^{\alpha \xi/b}~\frac{1}{n_k!}~\left\{V^{\prime}~
{\textstyle \phi_k (T) \,  
e^{\frac{ (\mu  - (\lambda - i\eta) bT )k}{T} }}\right\}^{n_k} \right] \Theta(V^\prime) = \nonumber \\
&&\hspace*{-0.1cm}\int_0^{\infty}\hspace*{-0.2cm}dV^{\prime}
\int\limits_{-\infty}^{+\infty} d \xi~ \int\limits_{-\infty}^{+\infty}
  \frac{d \eta}{\sqrt{2 \pi}} ~ { \textstyle e^{ i \eta (V^\prime - \xi) - \lambda V^{\prime} 
+ V^\prime {\cal F}(\xi, \lambda - i \eta) } }\,.
%
%
%
\end{eqnarray}
After changing the integration variable $V \rightarrow V^{\prime} = V - b \sum\limits_k^{\alpha \xi/b} k~n_k $,
the constraint of $\Theta$-function has disappeared.
Then all $n_k$ were summed independently leading to the exponential function.
Now the integration over $V^{\prime}$ in Eq.~(\ref{five})
can be straightforwardly done resulting in
\begin{equation}\label{six}
\hspace*{-0.4cm}\hat{\cal Z}(\lambda,T,\mu) = \int\limits_{-\infty}^{+\infty} \hspace*{-0.1cm} d \xi
\int\limits_{-\infty}^{+\infty} \hspace*{-0.1cm}
\frac{d \eta}{\sqrt{2 \pi}} ~ 
\frac{  \textstyle e^{ - i \eta \xi }  }{{\textstyle \lambda - i\eta ~-~{\cal F}(\xi,\lambda - i\eta)}}~,
\end{equation}

\vspace*{-0.3cm}

\noindent
where the function ${\cal F}(\xi,\tilde\lambda)$ is defined as follows 
\begin{eqnarray}\label{seven}
&&\hspace*{-0.4cm}{\cal F}(\xi,\tilde\lambda) = \sum\limits_{k=1}^{\alpha\xi/b } \phi_k (T) 
~e^{\frac{(\mu  - \tilde\lambda bT)k}{T} }
= \nonumber \\
&&\hspace*{-0.4cm}\left(\frac{m T }{2 \pi} \right)^{\frac{3}{2} } \hspace*{-0.1cm} \left[ z_1
~{\textstyle e^{ \frac{\mu- \tilde\lambda bT}{T} } } + \hspace*{-0.1cm} \sum_{k=2}^{\alpha\xi/b }
k^{-\tau} e^{ \frac{(\mu + W - \tilde\lambda bT)k - \sigma k^{2/3}}{T} }  \right]\,.\,
\end{eqnarray}

As usual, in order to find the GCP by  the inverse Laplace transformation,
it is necessary to study the structure of singularities of the isobaric partition (\ref{seven}). 

\vspace*{-0.1cm}

\section{Isobaric Partition Singularities}

\vspace*{-0.2cm}

The isobaric partition (\ref{seven}) of the CSMM is, of course, more complicated
than its SMM analog \cite{Bugaev:00,Bugaev:01}
because for finite volumes the structure of singularities in the CSMM 
is much richer than in the SMM, and they match in the limit $V \rightarrow \infty$ only.
To see this let us first make the inverse Laplace transform:
\begin{eqnarray}\label{eight}
&&\hspace*{-0.6cm}{\cal Z}(V,T,\mu)~ = 
\int\limits_{\chi - i\infty}^{\chi + i\infty}
\frac{ d \lambda}{2 \pi i} ~ \hat{\cal Z}(\lambda,  T, \mu)~ e^{\textstyle   \lambda \, V } =
\nonumber \\
&&\hspace*{-0.6cm}\int\limits_{-\infty}^{+\infty} \hspace*{-0.1cm} d \xi
\int\limits_{-\infty}^{+\infty} \hspace*{-0.1cm}  \frac{d \eta}{\sqrt{2 \pi}}  
\hspace*{-0.1cm} \int\limits_{\chi - i\infty}^{\chi + i\infty}
\hspace*{-0.1cm} \frac{ d \lambda}{2 \pi i}~ 
\frac{\textstyle e^{ \lambda \, V - i \eta \xi } }{{\textstyle \lambda - i\eta ~-~{\cal F}(\xi,\lambda - i\eta)}}~= 
\nonumber \\
&&\hspace*{-0.6cm}\int\limits_{-\infty}^{+\infty} \hspace*{-0.1cm} d \xi
\int\limits_{-\infty}^{+\infty} \hspace*{-0.1cm}  \frac{d \eta}{\sqrt{2 \pi}}
\,{\textstyle e^{  i \eta (V - \xi)  } } \hspace*{-0.1cm} \sum_{\{\lambda^*_n\}}
e^{\textstyle  \lambda^*_n\, V } 
\left[1 - \frac{\partial {\cal F}(\xi,\lambda^*_n)}{\partial \lambda^*_n} \right]^{-1} \hspace*{-0.2cm},
\end{eqnarray}
where the contour  $\lambda$-integral is reduced to the sum over the residues of all singular points
$ \lambda = \lambda^*_n + i \eta$ with $n = 1, 2,..$, since this  contour in complex $\lambda$-plane  obeys the
inequality $\chi > \max(Re \{  \lambda^*_n \})$.  
Now both remaining integrations in (\ref{eight}) can be done, and the GCP becomes 
\begin{equation}\label{nine}
{\cal Z}(V,T,\mu)~ = \sum_{\{\lambda^*_n\}}
e^{\textstyle  \lambda^*_n\, V }
\left[1 - \frac{\partial {\cal F}(V,\lambda^*_n)}{\partial \lambda^*_n} \right]^{-1} \,,
\end{equation}
i.e. the double integral in (\ref{eight}) simply  reduces to the substitution   $\xi \rightarrow V$ in
the sum over singularities. 
This is a remarkable result which 
can be formulated as the following 
\underline{\it theorem:}
{\it if the Laplace-Fourier image of the excluded volume GCP exists, then
for any additional $V$-dependence of ${\cal F}(V,\lambda^*_n)$ or $\phi_k(T)$
the GCP can be identically represented by Eq. (\ref{nine}).}

%
%
\begin{figure}[ht]
\includegraphics[width=8.6cm,height=6.0cm]{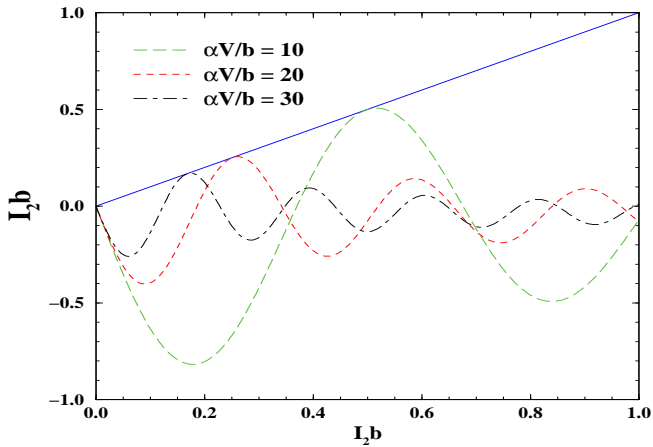}
\caption{A graphical solution of Eq. (\ref{twelve}) for $T = 10$ MeV and $\tau = 1.825$.   
The l.h.s. (straight line) and  r.h.s. of Eq. (\ref{twelve}) (all dashed curves) are shown 
as the function of
dimensionless parameter $I_2\,b$ for the three values of the largest fragment size $\alpha V/b$. 
The intersection point at $(0;\,0)$ corresponds to a real root of Eq. (\ref{ten}).  
Each tangent point with the straight line generates  two complex  roots of (\ref{ten}).
}
  \label{fig1}
\end{figure}



The simple poles in (\ref{eight}) are defined by the  equation 
\begin{equation}\label{ten}
\lambda^*_n~ = ~{\cal F}(V,\lambda^*_n)\,.
\end{equation}
In contrast to the usual SMM \cite{Bugaev:00,Bugaev:01} the singularities  $ \lambda^*_n $ 
are (i) functions of volume $V$,
and (ii) they can have a non-zero imaginary part, but 
in this case there  exist  pairs of complex conjugate roots of (\ref{ten}) because the GCP is real.

Introducing the real $R_n$ and imaginary $I_n$ parts of  $\lambda^*_n = R_n + i I_n$,
we can rewrite  Eq. (\ref{ten})
as a system of coupled transcendental equations ($K_{up} = \alpha V/b $)  
\begin{eqnarray}\label{eleven}
&&\hspace*{-0.2cm} R_n = ~ \sum\limits_{k=1}^{K_{up} } \tilde\phi_k (T)
~{\textstyle e^{\frac{Re(\nu)\,k}{T} } } \cos(I_n b k)\,,
\\
\label{twelve}
&&\hspace*{-0.2cm} I_n = - \sum\limits_{k=1}^{K_{up} } \tilde\phi_k (T)
%
%
~{\textstyle e^{\frac{Re(\nu)\,k}{T} } } \sin(I_n b k)\,,
\end{eqnarray}
where we have introduced the effective chemical potential $\nu = \mu + W (T)  - \lambda^*_n b\,T$, and 
the reduced distributions $\tilde\phi_1 (T) = z_1 \exp(-W(T)/T)$ and 
$\tilde\phi_{k > 1} (T) = k^{-\tau}\, \exp(-\sigma (T)~ k^{2/3}/T)$ for convenience.

%
%
 \begin{figure}[ht]
  \includegraphics[width=8.6cm,height=6.0cm]{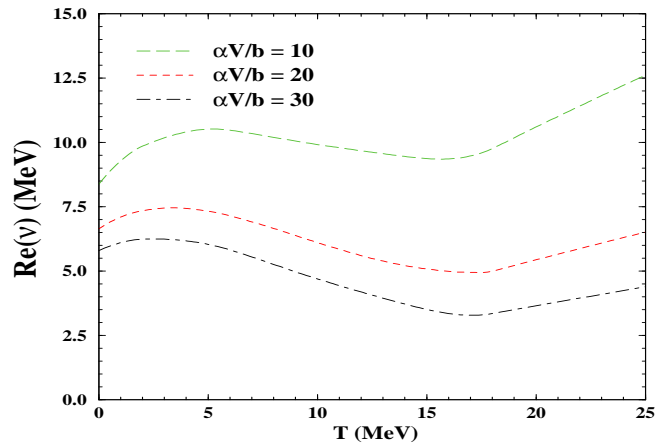}
  \caption{Each curve separates  the $T-Re(\nu)$ region of one real root of Eq. (\ref{ten})
(below the curve), three complex roots (on the curve) and five and more roots (above the curve)
for three values of $\alpha V/b$ and the same parameters as in Fig. 1.
}
  \label{fig2}
\end{figure}

Consider the real root $(R_1 > 0, I_1 = 0)$, first. 
For $I_n = I_1 = 0$ the real root $R_1$ exists for any $T$ and $\mu$.
Since for $I_n \neq 0$ defined by (\ref{twelve}) the inequality $\cos(I_n b k) \le 1$ cannot
become the equality for all values of $k$ simultaneously, then from Eq. (\ref{eleven})  
one obtains
\begin{equation}\label{thirteen}
R_n < \sum\limits_{k=1}^{K_{up} } \tilde\phi_k (T)
~{\textstyle e^{\frac{Re(\nu)\, k}{T} } } \quad \Rightarrow \quad R_n < R_1\,, 
\end{equation}
where the second inequality (\ref{thirteen}) immediately follows from the first one.
Note that the second inequality (\ref{thirteen}) plays a decisive role in the thermodynamic limit
$V \rightarrow \infty$ because in this case  it generates 
the pressure of the gaseous phase.  

Like in the usual SMM \cite{Bugaev:00,Bugaev:01}, 
for infinite volume the effective chemical potential must be real and non-positive, 
$\nu \le 0$,
because in the latter case the function ${\cal  F}(V,  \lambda_1^*)$ (\ref{seven}) diverges and 
the formal manipulations
in (\ref{five})  to establish (\ref{six}) cannot be used. 
The limiting value $\nu = 0$ defines the liquid phase singularity of the
isobaric partition which gives the liquid pressure 
$p_l(T,\mu) = R_1 T = (\mu + W(T))/b$ \cite{Bugaev:00,Bugaev:01}. 
But for finite volumes and finite $K_{up}$ the effective  chemical potential can
be complex (with either sign for its real part)  and its value defines the number and position of the imaginary roots 
$\{\lambda^*_{n > 1} \}$ 
in the complex plane.
Consider the natural values of $K_{up}$, first. 
As it is seen from  Fig. 1., the r.h.s. of (\ref{twelve}) is the amplitude and frequency modulated sine-like  
function of dimensionless parameter $I_n\,b$. 
Therefore, depending on $T$ and $Re(\nu)$ values there may exist either zero, 
or finite or infinite number of 
complex roots $\{\lambda^*_{n>1}\}$. In Fig. 1. we showed a special case which corresponds to  exactly three 
roots of Eq. (\ref{ten}) for each value of $K_{up}$: the real root ($I_1 = 0$) and two complex conjugate
roots ($\pm I_2$). 
Since 
the r.h.s. of (\ref{twelve}) is monotonously increasing function
of  $Re(\nu)$, when the former is positive,  
then it is possible to map the $T-Re(\nu)$ plane into
regions of a fixed number of roots of Eq. (\ref{ten}). 
Each curve in \mbox{Fig. 2.} divides the $T-Re(\nu)$ plane
into three parts: for $Re(\nu)$-values below the curve there  is only a real root, 
for points on  the curve there exist      
three roots, and above the curve there are five and more roots of Eq. (\ref{ten}).

A similar situation occurs  for the real values of $K_{up}$. In this case all sums in 
Eqs. (\ref{ten}-\ref{thirteen}) should be expressed via the Euler-MacLaurin   
formula
\begin{eqnarray} \label{fourteen}
&&\hspace*{-0.4cm}\sum\limits_{k=1}^{K_{up} } f_k = f(1) + \hspace*{-0.2cm} \int\limits_2^{K_{up}}
\hspace*{-0.15cm} dk\, f(k) + 
{\textstyle \frac{f(K_{up}) + f(2)}{2}  + \Delta_f (K_{up}) - }\,
 \nonumber\\
&&\hspace*{-0.4cm}\Delta_f (2)\,, {\rm where} \quad \Delta_f (K) = \sum\limits_{n=1} \frac{B_{2n} }{(2n)!} 
~\frac{d^{2n} f(x)}{d\,x^{2n} }\biggr|_{x = K}\,.
\end{eqnarray}
Here $B_{2n}$ are the Bernoulli numbers. 
The representation (\ref{fourteen}) 
allows one 
to study the effect of finite volume (FV) on the GCP (\ref{nine}). 

\vspace*{-0.1cm}

\section{Finite Volume Thermodynamics}

\vspace*{-0.2cm}

In the CSMM there are  two different ways of how the finite volume affects thermodynamical functions: 
for finite $V$ and $Re(\nu)$ there is always a finite number of simple poles in (\ref{nine}) and all of them
contribute into thermodynamic quantities; also the parameter $\alpha < 1$ describes a  difference 
between  the geometrical shape of the volume under consideration  and that one of the largest fragment
(assumed to be spherical).  
To see this, let us study the mechanical pressure which corresponds to the GCP (\ref{nine})

\vspace*{-0.5cm}

\begin{eqnarray} \label{fifteen}
&&\hspace*{-0.4cm}p
= T \frac{\partial \ln {\cal Z}(V,T,\mu) }{\partial V} = 
{\textstyle \frac{T}{{\cal Z}(V,T,\mu) } } \sum\limits_{\{\lambda^*_n\}} \Biggl[ 
\frac{\lambda^*_n~ e^{\lambda^*_n \,V} }{ 1 - \frac{\partial {\cal F}(V,\lambda^*_n)}{\partial \lambda^*_n}  } 
+   \nonumber\\
&&\hspace*{-0.4cm}\frac{  e^{\lambda^*_n \,V} }{\left[ 
1 - \frac{\partial {\cal F}(V,\lambda^*_n)}{\partial \lambda^*_n} \right]^2  }
\Biggl\{ b^2\,\frac{ \partial \lambda^*_n}{\partial V}  \sum\limits_{k=1}^{K_{up} } \tilde\phi_k (T)\,k^2
~{\textstyle e^{\frac{\nu\,k}{T} } }  + \tilde\phi_{K_{up}} (T) \times
\nonumber \\
&&\hspace*{-0.4cm}
{\textstyle e^{\frac{\nu\,K_{up}}{T} } } 
{\textstyle K_{up} 
\left[ (1 - \alpha) +  \frac{\nu}{T} \left( \frac{1}{2} - \alpha \right) \right]}+  
o(K_{up})\Biggr\} \Biggr]\,,
%
\end{eqnarray}

\vspace*{-0.3cm}

\noindent
where we give the main term for $\lambda^*_0$ and  leading FV corrections explicitly for $Re(\nu)/T < 1$, whereas  
$o(K_{up})$ accumulates the higher order corrections due to  
the  Euler-MacLaurin Eq. (\ref{fourteen}).
In evaluation of (\ref{fifteen}) we used an explicit representation of the
derivative $\partial \lambda^*_n/ \partial V$ which can be found from Eqs. (\ref{ten}) 
and (\ref{fourteen}). 
The first term in the r.h.s. of (\ref{fifteen}) describes the partial pressure generated
by the simple pole $\lambda^*_n $ weighted with the ``probability'' $e^{\lambda^*_n \,V }/{\cal Z} (V,T,\mu)$, 
whereas the second and third terms appear due to the volume dependence of $K_{up}$. 
Note that, instead of the FV corrections, the usage of  natural values for 
$K_{up}(V)$ would generate the artificial delta-function   
terms in (\ref{fifteen}) for the volume derivatives.

As one can see from (\ref{fifteen})
for finite volumes the corrections can give a non-negligible contribution to the pressure because
in this case 
$Re(\nu) > 0$ can be positive.   
The real parts of the partial pressures $T \lambda^*_n$  may have either sign. Therefore,
according  (\ref{thirteen}) 
the positive pressures $T R_n > 0$ are metastable and the negative ones $T R_n < 0$ are mechanically unstable. 
The pair of complex conjugate roots with the same value of $T R_n$ 
corresponds to a formation and decay of those states in thermodynamical system at finite volumes. 

When $V$ increases the number of simple poles in (\ref{eight}) also increases and imaginary part of 
the closest to the real $\lambda$-axis  poles becomes very small. 
For infinite volume the  infinite number of simple poles moves toward 
the real $\lambda$-axis to the vicinity of liquid phase singularity $p_l(T,\mu)/T$ and, thus,
generates  an essential singularity of function ${\cal F}(V, p_l/T)$ in (\ref{seven}). 
In this case the contribution of any of remote poles  from the real $\lambda$-axis  
to the GCP vanishes.  
Then it can be  shown that the FV corrections in (\ref{fifteen})
become negligible  because of the inequality $Re(\nu) \le 0$, and, consequently,
the reduced distribution of largest fragment 
$\tilde\phi_{K_{up}} (T) = K_{up}^{-\tau}\, \exp(-\sigma (T)~ K_{up}^{2/3}/T)$
and the
derivative $\partial \lambda^*_n/ \partial V$
vanish for all $T$-values, and we obtain the usual SMM solution \cite{Bugaev:00,Bugaev:01},
and its thermodynamics is governed by the farthest right singularity in complex $\lambda$-plane.
The corrections of a   similar kind should appear in the entropy, particle number and energy density
because of the $T$ and $\mu$ dependence of $\lambda^*_n$ due to (\ref{ten}) \cite{Bugaev:04}. 
Therefore, these corrections should be taking into account while analyzing the experimental yields
of fragments. Then the phase diagram of the nuclear liquid-gas phase transition
can be recovered from the experiments on finite systems (nuclei)  with more confidence.  

Also it is  possible  that the metastable and unstable modes can emerge in the dynamically expanding system
created in experiments, or even there may exist  a direct relation between these modes and the spinodal
instabilities discussed with respect to phase transition in finite systems \cite{Randrup:04},
but it is theme of the other work.

{\bf  Acknowledgments.}
This work was supported by the US Department of Energy.
The fruitful  discussions with Y. H. Chung and M. I. Gorenstein are appreciated.




\end{document}